\documentclass[sigconf]{acmart}

\copyrightyear{2025}
\acmYear{2025}
\setcopyright{cc}
\setcctype{by}
\acmConference[ICHEC 2025]{Proceedings of the 2025 International Conference on Human-Engaged Computing}{November 21--23, 2025}{Singapore, Singapore}
\acmBooktitle{Proceedings of the 2025 International Conference on Human-Engaged Computing (ICHEC 2025), November 21--23, 2025, Singapore, Singapore}
\acmDOI{10.1145/3786995.3787061}
\acmISBN{979-8-4007-2234-9/2025/11}

\AtBeginDocument{%
  }

\begin{document}

\title{From Meme to Method: Rethinking Animal Adoption Platforms through the Cat Distribution System}

\author{Carl Angelo Angcana}
\email{cgangcana@up.edu.ph}
\affiliation{
  \institution{Institute of Computer Science, University of the Philippines Los Baños}
  \city{Los Baños}
  \country{Philippines}
}

\author{Jamlech Iram Gojo Cruz}
\email{jngojocruz@up.edu.ph}
\orcid{0009-0001-8164-8778}
\affiliation{
  \institution{Institute of Computer Science, University of the Philippines Los Baños}
  \city{Los Baños}
  \country{Philippines}
}

\begin{abstract}
  The internet folklore of the Cat Distribution System (CDS) humorously suggests that cats are “assigned” to people rather than intentionally sought. Beyond its playful origins, CDS reflects a culturally resonant way people perceive and engage in adoption, and this user context can guide the redesign and improvement of adoption systems. In the Philippines, where an estimated 13.11 million stray cats and dogs place the country sixth worldwide in overpopulation, this framing offers a novel way to rethink adoption platforms. We developed a prototype application inspired by CDS principles, focusing on features such as algorithmic matchmaking, community reporting, and proximity-based discovery. An initial evaluation with potential users (n=35) indicated that the system was positively received for its ease of use and its alignment with users’ intuitive expectations, though participants highlighted areas for improvement in transparency of matchmaking and owner–adopter communication. The findings suggest that culturally embedded metaphors like CDS can shape mental models, making adoption processes feel more serendipitous and less transactional.
\end{abstract}




\begin{teaserfigure}
  \centering
  \includegraphics[width=\linewidth]{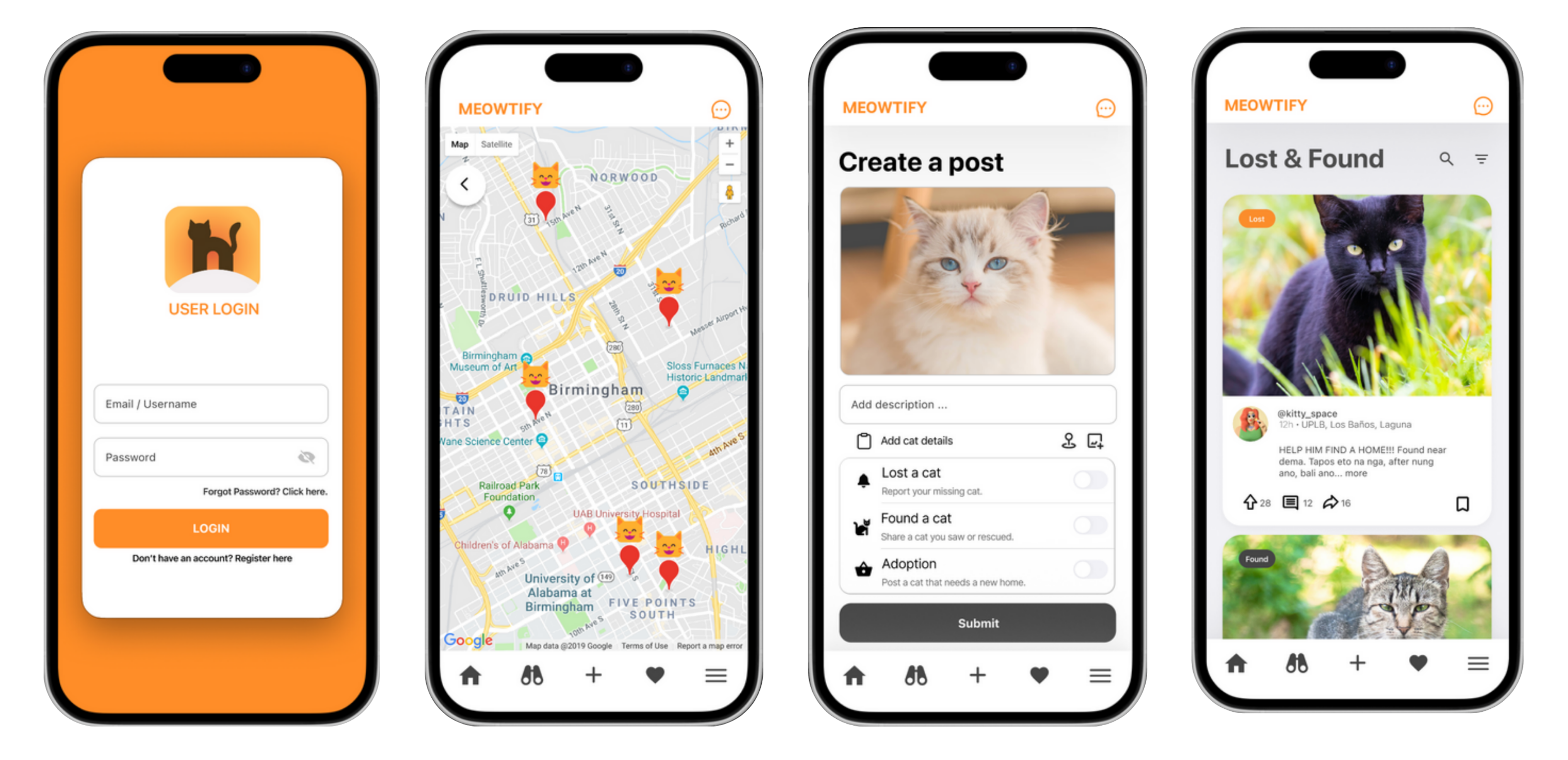}
  \caption{Sample Screenshots for Meowtify Organization of Cat Profile.}
  \label{fig:screenshot1}
\end{teaserfigure}

\maketitle

\section{Introduction}
Animal overpopulation remains a pressing welfare issue, with millions of stray cats and dogs struggling to find permanent homes. Despite the rise of online adoption platforms, adoption practices remain fragmented and heavily dependent on formal search-and-filter mechanisms. These systems prioritize efficiency and rational choice but often overlook the everyday dynamics that shape actual adoption outcomes, such as geographic proximity, the readiness and capacity of potential adopters, and the circulation of information within local communities.

In contrast, the internet meme of the Cat Distribution System (CDS) humorously frames adoption as serendipitous encounters in which cats “assign themselves” to people. While playful, CDS captures an alternative mental model of adoption that emphasizes proximity, chance, and social networks. This framing resonates strongly with cultural practices where adoption is less about systematic searching and more about community-driven encounters and informal decision-making. For human–computer interaction (HCI), CDS offers a culturally grounded perspective for rethinking how adoption platforms might better reflect the ways people perceive and engage with animal welfare.

We developed the Serendipitous Adoption framework, where we framed CDS as a mental model that can inform the design of digital adoption systems. By treating CDS not only as folklore but as an articulation of user expectations and behaviors, we explore how cultural narratives can be translated into design strategies. Specifically, our objectives are: 
\begin{enumerate}
    \item construct a CDS-informed mental model for adoption, 
    \item derive design features that align with this model, and 
    \item develop a prototype platform and conduct an initial evaluation with target users.
\end{enumerate}

The final CDS-inspired adoption system prototype was well-received, with participants commending its design idea and overall usability. This study contributes to HCI research and practice by demonstrating how culturally resonant narratives, even those originating from internet folklore, can guide the design of systems in meaningful ways. Beyond improving adoption outcomes, our approach highlights the value of grounding design in user mental models that emerge from culture and community, offering a pathway for designing systems that are both effective and contextually relevant.

\section{Background and Related Works}

\subsection{Cat Overpopulation and Limitations of Current Adoption Systems}
The overpopulation of stray and feral cats has become a pressing global issue, with significant implications for animal welfare, public health, and biodiversity. Globally, domestic cats are among the most widespread carnivores, with population growth fueled by high reproductive capacity and human abandonment. Estimates suggest that hundreds of millions of cats roam freely worldwide, with their numbers continuing to rise in urban and peri-urban areas where food and shelter are accessible. Cats are one of the most significant threats to biodiversity, killing between 1.3–4 billion birds and 6.3–22.3 billion mammals annually in the United States alone, and threatening endemic species in island and urban ecosystems \cite{medina2011global, loss2022review, loss2013impact}. Free-roaming cats are also reservoirs of zoonotic pathogens, such as rabies \cite{hatam2021toxoplasma, nguyen2021canine}. This poses significant public health risks, particularly in developing countries. The Philippines is among the top countries globally in terms of homeless animal overpopulation, with an estimated 13.11 million stray cats and dogs roaming the country, placing the Philippines  \cite{senate2024overpopulated}. 

Various approaches to population management have been implemented worldwide, with Trap-Neuter-Return (TNR) programs remaining the most widely adopted for stabilizing colonies \cite{levy2003evaluation, hostetler2020effective}. However, adoption has long been positioned as one of the most humane responses to overpopulation. Traditional shelter-based processes remain foundational, but research highlights persistent limitations. Animal characteristics, such as behavior, age, and appearance, strongly influence adoptability, while environmental factors can hinder the likelihood of adoption \cite{protopopova2012effects, powell2021characterizing}. The rise of online adoption platforms, such as Petfinder and Adopt-a-Pet, has expanded visibility and accessibility. Yet, these systems remain fragmented, with limited interoperability across shelters and rescues \cite{weiss2013community, horecka2022critical}. Human–computer interaction (HCI) research emphasizes that online adoption systems often overemphasize rational search tools while overlooking emotional and relational aspects of adoption, which can impact long-term outcomes \cite{slovak2023designing}. There are also recurring issues with returns to shelters due to mismatched expectations \cite{powell2021characterizing}. 

Social media has also become a dominant channel for adoption efforts. Campaigns on Facebook, Instagram, and TikTok increase visibility for adoptable animals and mobilize community support \cite{riddle2020social}. Though reliance on these introduces challenges such as unverified information and logistical mismatches \cite{applebaum2020concerns}.  Technology-mediated systems leverage recommender systems, mobile apps, and location-aware tools to enhance adoption by personalizing matches and fostering discovery \cite{bricman2023user, chang2023agapet}, yet these solutions remain underutilized.

\subsection{Mental Models, Cultural Frames, and Adoption Decision-Making}

Beyond functional and technological limitations, adoption is fundamentally shaped by human mental models. These are internal cognitive representations, such as informal understandings or folk theories, that users develop about how systems work. In turn, these shape their interactions and decision-making processes. In HCI, successful design emerges with users' mental models \cite{norman2013design, nielsen2024mental}.

The popular “Cat Distribution System” (CDS), a meme and piece of internet folklore that humorously suggests cats are “assigned” to humans rather than actively sought out, illustrates how cultural frames shape adoption. Although playful in origin, CDS reflects a broader folk theory: many people view adoption as a serendipitous encounter rather than a rational search process, and frame cats as agents of fate or chance \cite{zhang2024insights}. Internet memes function as cultural artifacts that express community values and shared understanding \cite{shifman2014memes, zhang2025memeing}. The CDS phenomenon proliferating on TikTok and other social media platforms represents a collectively constructed mental model of adoption as meaningful, emotionally resonant, and non-transactional.

These narratives align with empirical findings in the human-animal bond literature: adopters frequently describe pets as arriving at critical moments in life, which reinforces perceptions of purpose and meaning \cite{walsh2009human, scoresby2021pet}. Research also shows that adopters' preferences are often influenced by age, color, or behavior; those are considered folk theories of value that filter what is considered "adoptable" \cite{kogan2013cats}. These views are also supported by anthropological perspectives, which suggest that humans' tendency to attribute agency and providence to significant animal encounters \cite{prato2022complexity}. Recognizing the role of serendipity, folklore, and social media in shaping human decisions could lead to the design of adoption systems that align more closely with the lived mental models of adopters. This can also explain why purely rational systems remain fragmented while socially embedded practices continue to drive meaningful outcomes.

\subsection{Netnography and Culture-Inspired Design as Method for Rethinking Adoption Systems}
To understand and leverage these culturally embedded mental models, recent HCI research proposes culture-inspired design \cite{heimgartner2013reflections, heimgartner2017culturally}, which explicitly uses cultural elements, including memes and digital narratives, as sources of design inspiration and requirement specification. A study by Zhang demonstrated that memes serve as "emotional bridges" and reveal implicit expectations for platform design \cite{zhang2025memeing}. Similarly, the MEMEography method utilizes internet memes to gain an understanding of people and their contexts based on the memes they post in internet communities. This study establishes that memes function as empirical evidence of users' collective mental models \cite{kaltenhauser2021memeography}.

To systematically extract mental models from digital culture, netnography provides a rigorous methodological foundation. Netnography is a qualitative research methodology adapted from ethnography, specifically designed to study behaviours, cultures, and meanings within online communities through observation of digital artifacts \cite{kozinets2019netnography, kozinets2010netnography}. Here, qualitative coding techniques, including open, axial, and selective coding, are applied. The synthesis of findings is then used to discover design insights. This method, unlike traditional ethnography, enables researchers to observe naturalistic expressions of belief and meaning-making without disrupting communities \cite{bowler2010netnography}. It is also efficient in capturing unfiltered, context-rich expressions, making it particularly valuable for understanding culturally resonant phenomena, such as internet memes and CDS.

Netnographic observations gain rigor when combined with complementary methods, such as interviews or participatory workshops, also known as triangulation \cite{kozinets2010netnography}. The multi-method approach validates the interpretations from observations against explicit user articulations. This also ensures that the design decisions reflect both naturalistic behavior, as expressed in videos, and conscious user needs, as articulated in interviews.

\subsection{Aligning Adoption Platforms with User Mental Models}
By grounding adoption system design in users' mental models, like CDS, platforms can align with existing cognitive schemas rather than imposing purely transactional logic. Designing effective adoption platforms requires moving beyond rational search functions and toward supporting how people think and feel about the animal they will be adopting \cite{minnis2024decision, weiss2012did}.

Studies have shown that recommender systems that incorporate serendipity and affective elements increase engagement and satisfaction \cite{kotkov2016survey, zhang2012auralist}.  This suggests presenting pets not solely through rational matches (e.g., age, size, location), but also in ways that echo how people feel “chosen” by animals. Location-aware technologies, readiness assessments, and community reporting tools remain important, but their design must be ethically grounded by incorporating safeguards against abandonment, bias, and impulsive adoption through education and post-adoption support \cite{caraban201923, rahman2025trust}.

This study demonstrates how mental models derived from digital folklore can inform culturally resonant and ethically responsible design. Adoption systems can be designed to feel more intuitive, emotionally meaningful, and aligned with how communities actually conceptualize and engage in adoption practices.

\section{Methods}

This chapter outlines the research design, participants, instruments, and analytic procedures used to construct the mental model and eventually, the Serendipitous Adoption Framework to be used as the basis for prototype design.

\subsection{Research Design}

This study used a qualitative research approach, combining interviews and digital observation of social media content. The goal was to understand how people encounter stray cats both in person and online and reveal the social, cultural, and emotional meanings of adoption.

The study examined how unexpected but meaningful interactions, often described as the “Cat Distribution System” in online communities, shape a framework for serendipitous adoption. By integrating interview insights and social media narratives, the research explored how people interpret, share, and construct adoption stories through their lived experiences and online posts.

\subsection{Research Participants}

Eight (8) respondents participated in the study, selected through purposive sampling to capture a diverse range of perspectives related to cat care and welfare. The sample included educators, business owners, organization officers, animal shelter caretakers, and city veterinarians from Laguna, Batangas, and Manila. Purposive sampling was appropriate for this study because it prioritized participants with direct, relevant experience in adoption, rescue, and animal welfare. Given the exploratory nature of the research and the limited time available for data collection, a small but information-rich sample was sufficient to surface key themes and lived experiences.

\begin{table*}[htpb]
    \centering
    \caption{Participant Profiles and Cat-Related Involvement}
    \label{tab:participant_profiles}
    \begin{tabular}{p{2cm} p{4cm} p{2cm} p{7cm}}
    \toprule
    \textbf{Participant} & \textbf{Role/Occupation} & \textbf{Location} & \textbf{Cat Ownership / Involvement} \\
    \midrule

    P1 & Educator & Batangas & Owns 4 cats \\
    
    P2 & Educator & Manila & Owns 3 cats; rescued \\

    P3 & Business Owner & Cabuyao & Regular cat rescuer \\

    P4 & Officer of NGO (Cats of UPLB) & Los Ba\~nos & Fosters pets for adoption; manages social media adoptions \\

    P5 & City Veterinarian & Calamba & Oversees animal control and adoption \\

    P6 & City Veterinarian & Cabuyao & Oversees animal healthcare; manages FB page for lost-and-found animals \\

    P7 & Housewife / Entrepreneur & Cabuyao & Owns 12 cats; rescues and has surrender experience \\

    P8 & Animal Shelter Caretaker & Calamba & Works at animal shelter and dog pound facility; fosters pets \\
    
    \bottomrule
    \end{tabular}
\end{table*}

\subsection{Research Instruments}
This study employed three complementary research instruments to gather qualitative insights and prototype evaluation data:

\begin{enumerate}
    
        \item\textbf{Semi-Structured Interview Guide.}
            The primary instrument for understanding lived experiences was a semi-structured interview guide composed of open-ended questions focusing on participants’ interactions with stray cats. The interview explored:
            \begin{itemize}
                \item Personal experiences in adopting or rescuing cats
                \item Factors influencing adoption decisions
                \item Motivations and barriers encountered in the process
                \item The role of technology, online platforms, and social media
                \item Perceptions of the “Cat Distribution System” (CDS) concept
                \item Reflections on serendipitous encounters and how these shape human–cat relationships
            \end{itemize}
            This structure allowed participants to narrate their experiences in their own terms while enabling the researcher to explore shared meanings and emerging themes.
        \item\textbf{Observational Review Checklist for Social Media Content.}
        To complement interview data, the study used an observational checklist to analyze 30 publicly available Facebook posts and short-form videos depicting spontaneous or serendipitous encounters with stray cats. Each post was reviewed for key contextual and narrative indicators, such as:
        \begin{itemize}
            \item Location or setting of the encounter
            \item Type of interaction (e.g., stray approach, rescue, rehoming)
            \item Behavioral cues and emotional tone in the narration
            \item Indicators of cultural or moral interpretation (e.g., “swerte,” destiny, compassion, or obligation)
        \end{itemize}
        \item\textbf{Post-Task Questionnaire for Prototype Evaluation.} We used a post-task questionnaire as an evaluation method which contained the following:
        \begin{itemize}
            \item Likert-scale items adapted from the System Usability Scale (SUS) to capture perceptions of usability and confidence.
            \item Custom items targeting CDS-inspired design features (e.g., serendipity in matchmaking, usefulness of proximity, sense of community in reporting).
            \item Three open-ended questions asking participants what they liked most, what improvements they would suggest, and how they feel about the CDS-inspired features.
        \end{itemize}
\end{enumerate}

\subsection{Data Collection Procedure}
Data collection was conducted both online and in person between October and November 2025.

\begin{enumerate}
    \item\textbf{Phase 1 - Interview Data.}
    Semi-structured interviews were conducted with eight participants who had experience in rescuing, adopting, or caring for stray cats. Sessions lasted between 20 to 45 minutes and were facilitated either through in-person meetings or via online platforms. With the participants’ consent, all interviews were audio-recorded and later transcribed verbatim. To ensure confidentiality, identifiable details were removed and transcripts were anonymized before analysis.
    \item\textbf{Phase 2 - Observational Review of Short-Form Videos}
    To complement the interviews, 30 short-form videos were purposively selected from TikTok, Facebook, and Instagram. These videos were publicly accessible and featured spontaneous interactions between humans and stray cats. Sampling criteria included.
    \begin{itemize}
        \item relevance to serendipitous or unplanned encounters,
        \item display of physical proximity or repeated appearances of a stray,
        \item rescue or rehoming situations, and
        \item observable moral reflection or emotional interpretation (e.g., compassion, perceived “swerte (fortune),” destiny, humor).
    \end{itemize}
    Each video was reviewed using an observational checklist to assess narrative content, emotional framing, human behavior toward the animal, and cultural expressions embedded in the post. These observations served as naturally occurring examples of how people interpret stray encounters in everyday life.
\end{enumerate}

Before collecting any data, the researchers invited participants to join voluntarily and obtained their informed consent. Participants learned about the study’s goals, how their data would be used, and how their privacy would be protected. Audio recordings and transcripts were kept secure and will be deleted after the study ends. The report does not include any identifying details, and pseudonyms were used for everyone.

Only publicly accessible content was reviewed, ensuring that no private or restricted posts were included in the analysis. To protect user privacy, screenshots, profile names, captions, and any personally identifiable information were excluded from the research report. All observational notes were stored securely and will be deleted upon completion of the study, in accordance with the project’s data management protocol.

\subsection{Data Collection Analysis}
Data from the interviews and video observations were analyzed using thematic analysis, following the six-phase process \cite{braun2006using}. Both data sources were integrated to strengthen the depth and validity of the findings.
\begin{enumerate}
    \item\textbf{Familiarization:} The researcher read and re-read all interview transcripts and reviewed the observational notes from the short-form videos to gain an initial understanding of the data. Preliminary reflections were documented to guide early impressions.
    \item\textbf{Observation and Initial Coding:}
    Key meanings were coded across both sources, focusing on instances of chance encounters, care instincts, emotional responses, and cultural interpretations. Codes were applied consistently to capture patterns related to proximity, serendipity, environmental cues, and moral or practical decision-making.
    \item\textbf{Theme Identification:}
    Codes were grouped into potential themes based on conceptual similarity. Four overarching thematic categories emerged: 
    \begin{itemize}
        \item Serendipity
        \item Proximity
        \item Environment
        \item Decision
    \end{itemize}
    These themes reflect how individuals understand, interpret, and engage with stray cats in everyday contexts.
    \item \textbf{Theme Review and Refinement:}
    Themes were reviewed in relation to the coded data, interview narratives, and video observations to ensure internal coherence and conceptual clarity. Cross-comparison between interviews and social media observations strengthened the consistency and robustness of the themes.
    \item \textbf{Themes Definition and Meaning: }
    Themes were formally defined and named, ensuring they captured the essence of participants’ experiences and the broader cultural meanings expressed in online posts.
    \item \textbf{Framework Development:}
    The finalized themes were used to develop the Serendipitous Adoption Framework (Figure~\ref{fig:saf}), which illustrates how people make sense of adoption across both physical and digital contexts.
\end{enumerate}

\section{Findings}

This section presents the findings from semi-structured interviews with eight participants (Pn): educators, rescuers, organization officers, animal shelter staff, and city veterinarians from Batangas, Manila, and select municipalities of Laguna (i.e., Cabuyao, Calamba, and Los Baños). The interviews revealed four themes: serendipity, proximity, environment, and decision-making. These themes show how people encounter stray cats and how they interpret these encounters in their homes, workplaces, and communities.

\subsection{Thematic Analysis of Interviews}

\paragraph{\textbf{Theme 1: Serendipity as the Starting Point of Adoption}}

Among participants, adoption frequently resulted from unplanned encounters that gradually acquired significance. This pattern aligns with the central premise of the Cat Distribution System folklore, which posits that cats “choose” individuals through spontaneous appearances.

\begin{itemize}
    \item P1 recounted an instance in which a stray cat ``just showed up'' and, over time, became integrated into the household.
    \item P2 observed that cats ``simply appear,'' prompting humans to respond by providing food or care.
    \item P3 highlighted the instinctive impulse to rescue when encountering vulnerable kittens in public spaces.
    \item P4 and several other respondents acknowledged a sense of ``connection'' or emotional resonance that develops prior to formal adoption.
\end{itemize}

Taken together, these stories suggest that adoption often begins with a chance encounter, aligning with the ``serendipity'' dimension of the model. The unpredictability of these encounters fosters emotional investment, which later shapes behavior, commitment, and willingness to adopt.

\paragraph{\textbf{Theme 2: Proximity and ``Everyday Encounter'' as Catalysts for Adoption}}

Participants often linked adoption to being physically close, whether at home, in neighborhoods, or at work. Being nearby helped build trust and made caregivers more responsive.

\begin{itemize}
    \item Respondents said they met cats at their doorsteps, at work, or while commuting.
    \item Being close led to familiarity, more frequent interactions, and a sense of responsibility, as shown by comments like ``they keep coming back'' and ``they always appear.''
    \item Respondents noted that being nearby makes it easier to see a cat’s personality, friendliness, and approachability, which all affect adoption choices.
\end{itemize}

Proximity acts as both a practical factor, by making cats more visible and accessible, and as an emotional trigger that helps build rapport. This matches the prototype’s design, which is based on location and proximity, and shows it fits well with participants’ real experiences and the system’s goals.

\paragraph{\textbf{Theme 3: Environmental Conditions and Cultural Realities Shape Adoption Behavior}}

The interviews highlight how structural, cultural, and community-level factors determine how people perceive and engage with adoption systems.

\textbf{Lack of Formal Systems and Organizational Constraints.}
City veterinarians described limited manpower, absence of structured rescue processes, and reliance on walk-in events, barangay support, or partner shelters.

\begin{itemize}
    \item P5 and P6 explained that adoption systems are constrained by resources, staffing, budgets, and lack of dedicated facilities.
    \item Shelters may be forced to rely on euthanasia when unclaimed animals stay too long.
\end{itemize}

\textbf{Cultural Beliefs and Collective Attitudes.} 
Participants, like P7, referenced:
\begin{itemize}
    \item “Cats bring good luck.”
    \item Low public appreciation for non-purebred cats
    \item Generational differences in attitudes toward adopting strays
\end{itemize}
These cultural frames shape how people assign meaning to stray cats, influencing whether encounters result in empathy, avoidance, or adoption.

\textbf{Information Gaps and Visibility Issues} Several participants, including P8, noted that shelters, LGUs, and groups have inconsistent presence online and offline:

\begin{itemize}
    \item Difficulty locating adoption centers
    \item Mixed content feeds (adoption + marketing posts)
    \item Limited updates on bulletin boards
    \item Difficulty tracking adoptable animals
\end{itemize} 
This confirms the environment theme: adoption is mediated by information infrastructures, physical spaces, and institutional practices—which are currently fragmented or insufficient.

\paragraph{\textbf{Theme 4: Decision-Making Informed by Health, Capacity, and Practical Constraints}}
Even though many encounters with stray cats happen by chance, the actual decision to adopt is not spontaneous. Participants explained that they still weigh their own capacity, the risks, and the responsibilities that come with caring for an animal. From their responses, several practical considerations stood out:

\textbf{What people check before deciding to adopt:}
\begin{itemize}
    \item the cat’s health or medical background (vaccines, condition, age);
    \item whether they can afford regular food and vet care;
    \item if the cat will get along with pets already at home;
    \item whether they have enough space to keep another animal;
    \item if they have time and energy to commit to long-term care.
\end{itemize}

\textbf{Common barriers:}
\begin{itemize}
    \item worry about hidden illnesses or unexpected medical expenses;
    \item the overall cost of raising a pet;
    \item confusion about where or how to start the adoption process;
    \item uncertainty about the reliability of online adoption posts;
    \item preference for purebred cats in some households;
    \item lack of systems that ensure accountability between adopters and rescuers.
\end{itemize}
Overall, the interviews show that people want to feel capable and prepared before adopting a cat. This suggests that adoption systems should make important information easier to access and trust, so potential adopters can make informed and confident decisions.

\paragraph{\textbf{Theme 5: Technology as a Tool for Visibility, Accountability, and Humane Practice}}
Participants viewed technology as a crucial tool for strengthening adoption systems particularly in the Philippines where information system processes are fragmented.

\textbf{Key technological roles identified:}

\begin{itemize}
    \item Increasing visibility of adoptable animals and updates
    \item Streamlining processes (profiling, claiming, rehoming, rescue requests)
    \item Providing centralized information (vaccination, history, location)
    \item Promoting responsible ownership (education, reminders, guidelines)
    \item Establishing accountability (tracking adopters, ensuring compliance)
    \item Supporting humane practices (reducing euthanasia through rehoming networks)
\end{itemize}

\textbf{Participants imagined a system that:}
\begin{itemize}
    \item It is organized, searchable, and clutter-free.
    \item May include media components such as photos, profiles, and the location of each animal.
    \item Can supports chat functions, screening tools, and verification.
    \item Can promote LGU-to-community partnerships.
\end{itemize}

This expectation directly validates the CDS prototype’s design: a system that formalizes serendipity, improves navigation, and reduces friction in adopters’ decision-making.

\subsection{Findings from Video Observations}
\textbf{Types of encounters:}
Analysis of the 30-video dataset from Facebook, Instagram, and TikTok revealed two primary types of human–cat encounters: (1) rescue situations and (2) serendipitous approaches. These patterns show how unplanned interactions shape human perceptions of being chosen, feeling morally responsible, and deciding to adopt or care for stray cats.
\begin{enumerate}
    \item \textbf{Rescue Situations} 
    A significant portion of the videos show humans responding to a cat in distress. In these encounters, the cat’s safety or survival is visibly compromised, and viewers typically intervene. The common rescue scenarios from the dataset are:
    \begin{itemize}
        \item Cats left outside in the rain (Videos 2, 10)
        \item Injured strays follow rescuers, seeking help (Videos 3, 27)
        \item Kittens trapped in dangerous locations (roof, roadside, highway island) (Videos 12, 16, 26, 29)
        \item Cats scavenging through trash or showing signs of neglect (Videos 6, 28)
    \end{itemize}
    Individuals frequently provided immediate care, such as cleaning wounds, picking up the cat, feeding it, transporting it to a veterinarian, or offering shelter. In most rescue scenarios, they chose to adopt the cat during the initial encounter, showing how urgent situations trigger strong moral and emotional responses.

    \begin{figure*}[htbp] 
      \centering
      \includegraphics[width=0.5\linewidth]{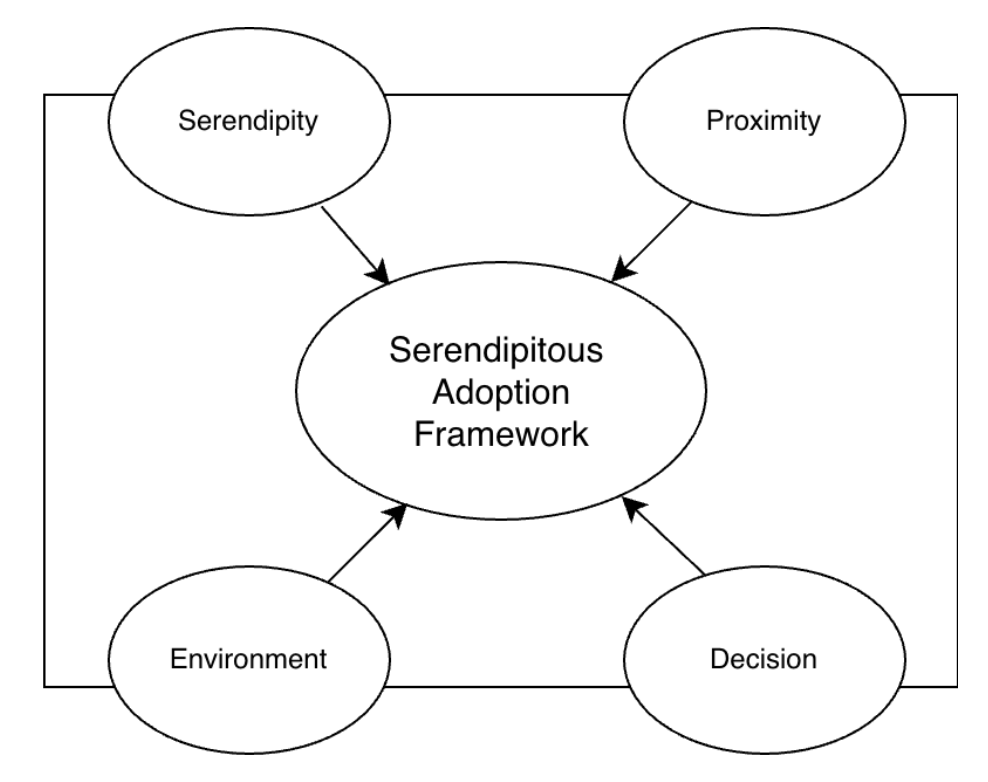}
      \caption{Serendipitous Adoption Framework Inspired by the Cat Distribution System (CDS) Folklore.}
      \Description{A simple diagram shows a central oval labeled “Serendipitous Adoption Framework,” surrounded by four ovals for Serendipity, Proximity, Environment, and Decision. Arrows from each oval point to the center, and lines connect the pairs Serendipity–Proximity and Environment–Decision to show they are related.}
      \label{fig:saf}
    \end{figure*}
    

    \begin{table*}[htpb]
    \centering
    \caption{Summary of Findings Compared to the Conceptual Framework}
    \label{tab:framework_summary}
    \begin{tabular}{p{3.5cm} p{11.5cm}}
    \toprule
    \textbf{Framework Dimension} & \textbf{How It Appeared in Participant Data} \\
    \midrule

    \textbf{Serendipity} & Cats appear unexpectedly; adoption begins from chance encounters. \\

    \textbf{Proximity} & Adoption is made possible by physical closeness and repeated exposure. \\

    \textbf{Environment} & Cultural beliefs, LGU structures, information gaps, and community norms shape encounters. \\

    \textbf{Decision} & Adoption depends on capacity, health concerns, emotional readiness, and practical constraints. \\
    
    \bottomrule
    \end{tabular}
    \label{tab:converge}
\end{table*}

    \item \textbf{Serendipitous Approaches}
    Numerous videos show cats voluntarily approaching individuals, sometimes repeatedly which viewers interpret as the cat “choosing” the human (Videos 1, 4, 8, 9). These encounters appear spontaneous rather than orchestrated, reinforcing the concept of the “Cat Distribution System.” The patterns of serendipitous approach found in the dataset are the following:
    \begin{itemize}
        \item Cats repeatedly visiting someone’s home (Videos 4, 9)
        \item Strays following a person during travel or daily routines (Videos 1, 14, 15, 18, 20, 22)
        \item Cats initiating physical closeness, such as climbing up to a person (Videos 17, 24)
        \item Parent cats bringing their kittens to a trusted human (Video 25)
    \end{itemize}

    These videos frequently show emotional responses such as surprise, joy, comfort, and a sense of being “chosen.” Captions often highlight themes of luck, destiny, or cosmic selection. Notable language cues include:

    \begin{itemize}
        \item “Naghahanap ng amo” (Looking for an owner)
        \item “Dito ka na sa amin titira” (You’ll be with us)
        \item “The cat loves you.”
        \item “Is this the cat distribution system selecting us?”
        \item “Cat distribution system at work again”
    \end{itemize}

    We also identified a distinct behavioral pattern in serendipitous approaches. In reviewing the videos, we observed the following:
    \begin{itemize}
        \item Most serendipitous approaches involved repeated proximity, as cats returned to the same individual or location despite the absence of a prior relationship.
        \item Human behavior frequently shifted from passive observation to active decision-making, such as feeding, permitting the cat indoors, or formalizing adoption.
    \end{itemize}

    Analysis of the 30-video dataset reveals two recurring pathways to serendipitous adoption. Both pathways show how everyday encounters, whether dramatic or subtle, prompt humans to interpret these moments through themes of moral reflection, fate, empathy, and responsibility.

\end{enumerate}

\subsection{Development of the Mental Model}

Based on the results of the thematic analysis of eight participant interviews and thirty ethnographic social media videos, a conceptual model was constructed to illustrate the process of serendipitous adoption. The resulting Serendipitous Adoption Framework, as shown in Figure~\ref{fig:saf}, represents a user mental model derived from participants’ interpretations of their experiences with stray cats.

The mental model illustrates how the core ideas of the “Cat Distribution System” are reframed into a conceptual structure that explains adoption behavior as a product of four interacting dimensions: \textit{serendipity}, \textit{proximity}, \textit{environment}, and \textit{decision}. It begins with a stray encounter that prompts moral reflection, leading to the formation of emotional bonds and building of trust between humans and cats. These interactions are reinforced by trust, social validation and a sense of destiny, culminating in adoption.

Participants’ stories and insights provide empirical support for the proposed framework, demonstrating that adoption behavior in the Philippines is influenced by spontaneous encounters, environmental context, real-world constraints, and moral-practical considerations, which align with the components identified in the conceptual model. The findings converge strongly with the conceptual model, as shown in Table~\ref{tab:converge}.

\begin{figure*}
        [!htbp] 
      \centering
      \includegraphics[width=0.9\linewidth]{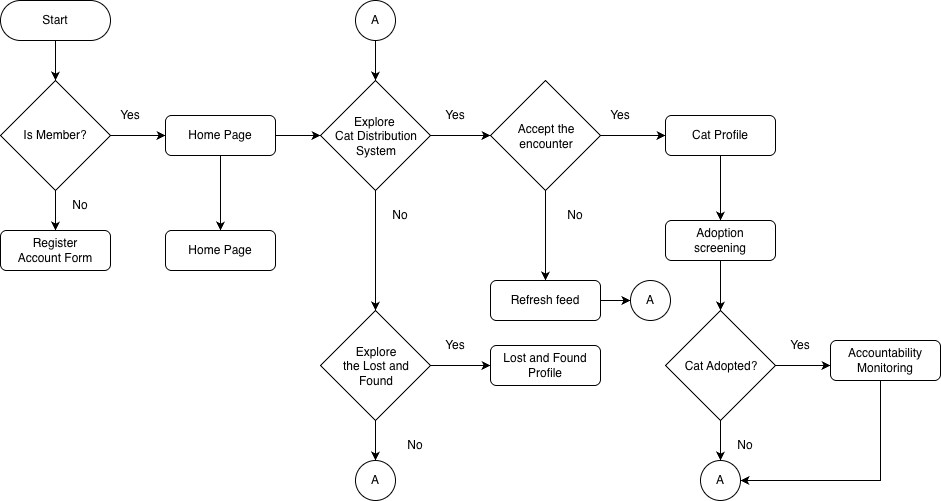}
      \caption{User Flow Diagram of the Meowtify Prototype based on the Cat Distribution System Folklore}
      \label{fig:userflow}
      \Description{Flowchart of the Meowtify user journey. Users start by checking if they are members and, if not, register an account, then reach the home page. From there they can explore the Cat Distribution System or the Lost and Found section. If they accept an encounter, they view a cat profile, go through adoption screening, and, if the cat is adopted, proceed to accountability monitoring; otherwise the flow returns to browsing.}
    \end{figure*}

This mental model enables users to interpret adoption opportunities as emergent rather than strictly planned. For example, some users perceive distribution as a sequence of serendipitous encounters, while others depend on environmental influences or proximity cues. Consequently, user actions such as browsing available cats, signaling interest, or committing to adoption are grounded in their mental representation of the CDS. The design of the CDS therefore aims to align the system’s conceptual model, or prototype, with the user’s mental model, ensuring coherence between how the system is structured or how users naturally understand the adoption process.

This model also reflects the shared mental schema found in digital narratives such as posts, captions, and comments that frame cat encounters as serendipitous, destined, or morally significant.

\section{Design Methods}
\subsection{User Flow and System Design}
    The initial concepts of the system were developed as low-fidelity sketches and wireframes using Figma. These early representations emphasized the structural layout of the Cat Distribution System (CDS), allowing the research team to define navigation flows, core functions, and content placement without prioritizing visual design. The low-fidelity wireframe phase facilitated rapid iteration and refinement of the conceptual model of the system prior to detailed interface design.

    Following the low-fidelity stage, high-fidelity wireframes were developed in Figma to incorporate visual design elements such as typography, color schemes, and interactive components. These wireframes closely represented the intended final system and provided a realistic depiction of the user interface. Subsequently, user flows were developed to simulate the system experience for the Meowtify Prototype, as described in Figure~\ref{fig:userflow}.

    This diagram illustrates the intended user journey across key system features, including membership verification, exploration modules, serendipitous encounters, lost and found modules, adoption screening, and accountability monitoring.

The proposed system incorporates several core features designed to facilitate adoption and enhance community engagement, mimicking the "Cat Distribution System" concept. The primary components are as follows:

\begin{enumerate}
    \item\textbf{Automated Matchmaking - }
    The system employs an algorithm to match cats with users, inspired by the idea that “the cat chooses the user.” It recommends cats based on geographic proximity and a degree of randomness, while still allowing users to specify preferences to improve match quality. Users may save a recommended cat to initiate the adoption process or skip to receive alternative suggestions.
    \item \textbf{Reporting Functionality - }
    The system enables users to report and access information regarding lost and found cats, thereby supporting timely community updates. This feature also allows users to flag emergencies or rescue situations, strengthening animal welfare efforts.
    \item \textbf{Proximity-Enhanced Features - }
    The system integrates geolocation and mapping tools to enhance matchmaking and reporting. These allow users to find adoption opportunities near their location, connect with nearby cats, and view an interactive map of cats available in their area.
    \item \textbf{Post Creation - }
    The system enables users to create posts across multiple categories, including cat adoption listings and lost or found cat reports. Adoption-related posts are integrated with the matchmaking component, whereas lost-and-found posts facilitate the reporting function. This structure enhances information dissemination and fosters collective community engagement.
    \item \textbf{Community Engagement - } The system promotes interaction and trust through integrated chat and comment functionalities. The chat feature allows adopters and cat owners to communicate directly, clarify adoption details, and coordinate meetings. Comment sections on posts enable community members to provide feedback, share experiences, and offer updates. Collectively, these tools enhance transparency, facilitate collaboration, and reinforce shared responsibility for animal welfare.
\end{enumerate}

\subsection{Prototype Design}
We developed the prototype as a mobile application grounded in the Cat Distribution System (CDS) metaphor. Building on the core features outlined in the methods section, this section details the system’s functional components. We evaluated user performance by measuring task completion and task duration, generating insights into the prototype’s functional effectiveness and the experiential value of its CDS-inspired design.

The prototype enables users to create posts, manage lost-and-found listings, and utilize an interactive map to locate nearby cats. Developed in Figma, it incorporates reusable interface components to demonstrate the application's intended functionality and draws conceptual inspiration from the folklore-based Cat Distribution System.

The application supports two distinct types of posts. The first, the lost-and-found post, is intentionally excluded from the automated distribution algorithm to prevent misidentification of stray cats and misrepresentation of owned pets. The second, the adoption post, is created when a caretaker temporarily houses a cat to facilitate rehoming. For adoption posts, the automated distribution system assigns a cat to a potential adopter, facilitates communication between the parties, and provides direct access to information about the cat’s current location.

Users can customize post content based on available details, such as the cat's physical characteristics. Health-related information may differ, since some cats require additional veterinary assessment before medical data can be confirmed. For lost-cat posts, owners may provide the animal’s last known location, which the system uses to recalibrate the distribution algorithm and enhance visibility among nearby users.

Meowtify functions as a platform for disseminating cat-related information and assigns multiple roles to its users. It broadens participation in animal welfare by enabling community members, in addition to formal rescue groups, to contribute to visibility, reporting, and potential rehoming efforts.

\begin{figure*}[!htbp] 
  \centering
  \includegraphics[width=0.95\linewidth]{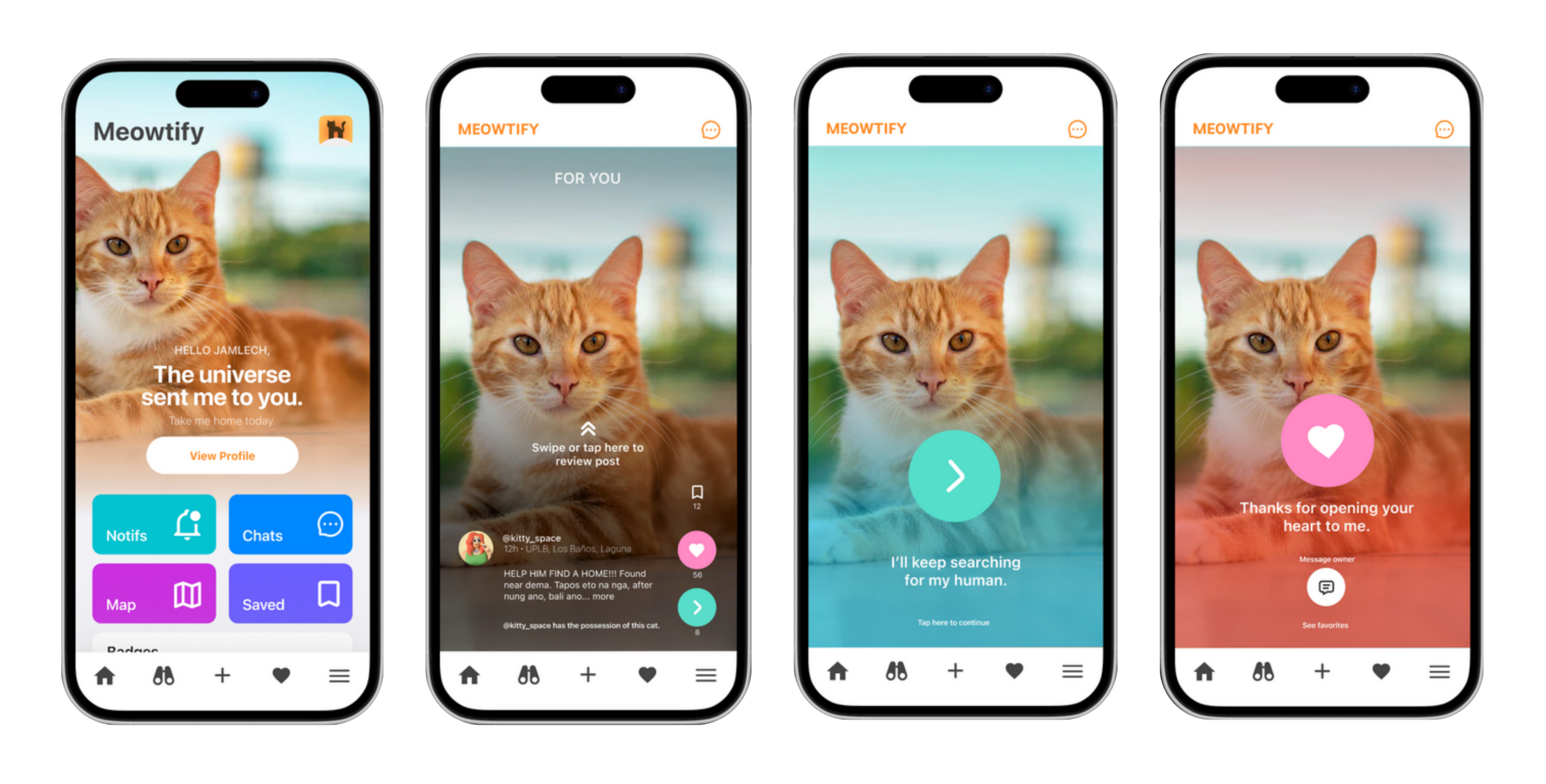}
  \caption{Sample Screenshots of Meowtify Exploration of Cat Distribution System}
  \label{fig:screenshot2}
  \Description{Four mobile screenshots of the Meowtify prototype.     The first screen shows a login page with a cat logo and fields for email or username and password. The second shows a city map with cat icons and pins marking cat locations. The third is a “Create a post” page with a cat photo, description box, and options for Lost a cat, Found a cat, and Adoption. The fourth shows a Lost & Found feed with cat photos, short text posts, and icons for likes and comments.}
    
\end{figure*}

Figure~\ref{fig:screenshot2} illustrates the portion of the prototype that simulates the automated “distribution” process inspired by the Cat Distribution System (CDS). When users access this screen, the system displays the message “The universe sent me to you,” indicating that a cat has been algorithmically matched to them. This interaction reflects the playful premise of CDS, in which it appears as though the cat selects the person rather than the user actively seeking the cat.

From this screen, users can access the cat’s profile, which includes community comments and reactions. These posts function as informal references that may shape the user’s perception of the match.

If the user selects the “love” button, the system displays a brief acknowledgement. This action does not finalize the adoption; instead, it enables the user to message the original poster through the built-in chat feature shown in Figure 4. If the user chooses to skip the match, the system simply generates another possible cat assignment.

\begin{figure*}[!htbp] 
  \centering
  \includegraphics[width=0.7\linewidth]{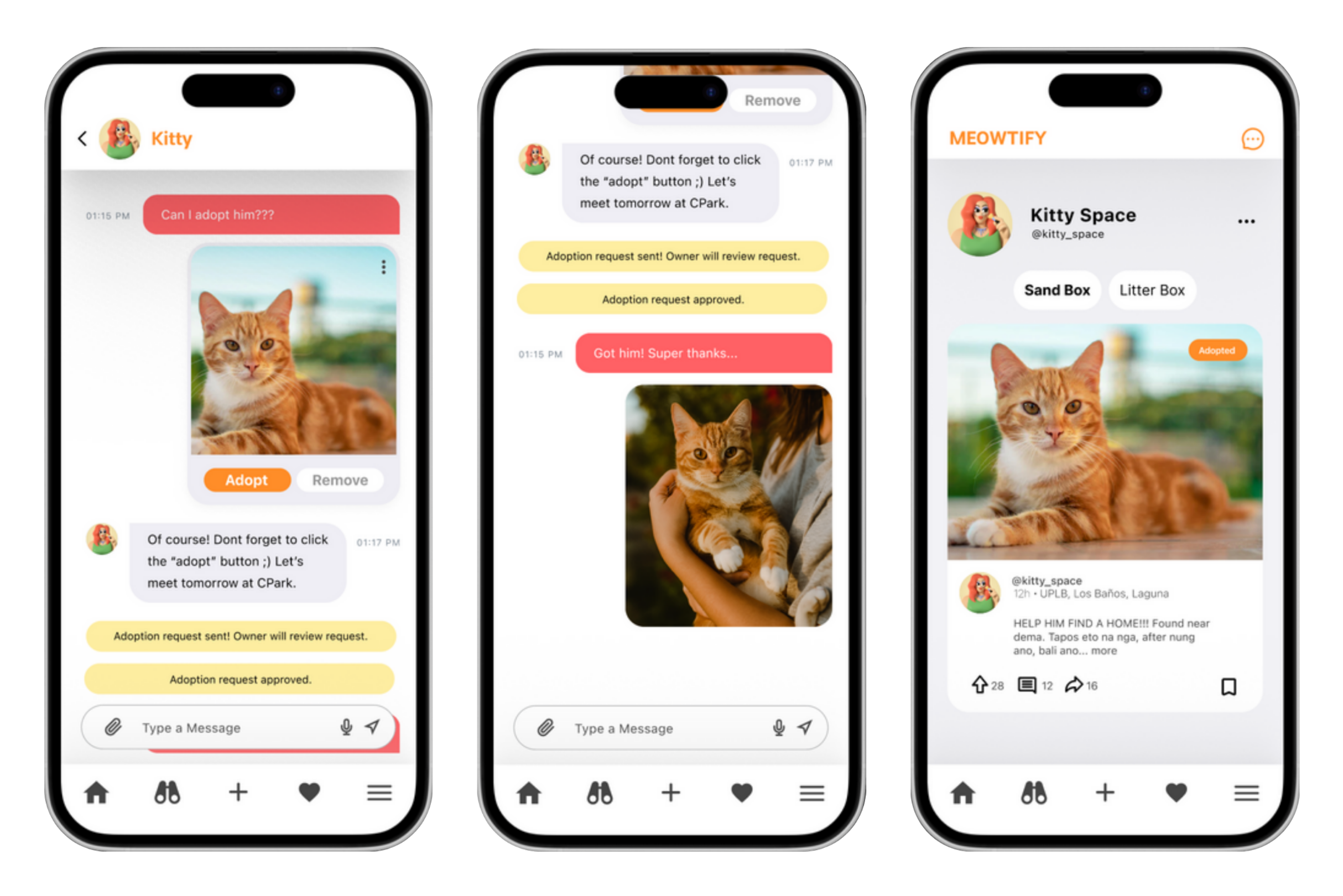}
  \caption{Screenshots of the Meowtify Adoption Feature. (Left) User interaction expressing intent to adopt a cat; (Middle) System confirmation of adoption approval; (Right) User profile displaying active posts in the “Sand Box” and archived posts in the “Litter Box.”}
  \Description{Three Meowtify screens showing the adoption conversation flow. The first two screens display a chat between a user and a cat owner, with photos of the cat, an “Adopt” button, and status messages such as “Adoption request sent” and “approved.” The third screen shows the cat’s post in the user’s profile area, marked as “Adopted” under the Sand Box and Litter Box tabs.}
  \label{fig:screenshot3}
\end{figure*}

Figure~\ref{fig:screenshot3} shows the messaging and adoption feature, which allows users to directly contact the person who posted the cat. Through this interface, potential adopters and caretakers exchange information, clarify details, and coordinate next steps, making the adoption process more organized and transparent.

The right-hand side of the interface includes two tabs: Sandbox and Litter Box. The Sandbox displays the user’s active posts that remain publicly visible, while the Litter Box stores posts that the system has already archived. When a cat is successfully adopted, the system automatically moves its post to the Litter Box. This process keeps the information up to date and prevents users from encountering cats that are no longer available.

Accurately maintained posts build user trust and support animal welfare. By avoiding outdated or misleading posts such as listings for cats that have already been adopted—the system reduces confusion and lowers the risk of fraudulent or mistaken transactions.

\subsection{Evaluation Design}

\subsubsection{\textbf{Selecting Participants}}
We recruited a small group of participants (N = 35) to try out the prototype across different evaluation sessions. In Stage 1, 23 participants completed a set of scenario-based tasks using the Maze-based setup and then answered the post-task questionnaire, which forms the sample for the quantitative usability ratings reported in Section 4.5. The group included individuals with and without prior pet adoption experience to ensure a diverse range of user perspectives during the usability test. A sample of this size is appropriate for early-stage usability evaluations, where the goal is to identify interface issues, user impressions, and practical concerns rather than to produce statistically generalizable results. Limited guidance was provided through video calls or in-person assistance to ensure smooth task completion. All participants provided voluntary informed consent.

\subsubsection{\textbf{Procedure}}
Participants were given access to the prototype and asked to complete a set of tasks:
\begin{itemize}
    \item Log-in to the app
    \item Explore matchmaking
    \item Adopt a cat
    \item Explore lost and found page
    \item Create a post
    \item Discover user profile
\end{itemize}

They were informed that the system was an early-stage prototype, not a fully functional application, and were encouraged to focus on the design features rather than app completeness. During the session, participants explored the prototype independently, then completed an online questionnaire.

\subsection{Data Analysis}
Quantitative data collected from Likert-scale items were analyzed descriptively to present an overview of usability and feature experience. Thematic analysis of open-ended responses was conducted to identify positive impressions, challenges, and recommendations for improvement.

\section{System Evaluation Results}

\subsection{Usability Testing: Quantitative Findings}

The quantitative evaluation examines how participants rated both the general usability of the prototype and the usefulness of its key features. We collected responses on a five-point Likert scale. To ensure clarity and avoid repeating long item descriptions, we refer to each statement by item number (Items 1–8). Table~\ref{tab:item_mapping} presents the full wording of the questionnaire items.

\begin{table}[h]
    \centering
    \caption{Mapping of item numbers to questionnaire statements.}
    \label{tab:item_mapping}
    \begin{tabular}{p{1cm} p{7cm}}
    \toprule
    \textbf{Item} & \textbf{Questionnaire Statement} \\
    \midrule
    Item 1 & I found the prototype easy to use. \\
    Item 2 & I felt confident using the prototype. \\
    Item 3 & The features of the prototype were well integrated. \\
    Item 4 & I would like to use a system like this in the future. \\
    Item 5 & The matchmaking feature (cats “sent” to me) captured the sense of chance or serendipity. \\
    Item 6 & The proximity/location/map aspect (cats near me) was useful. \\
    Item 7 & The reporting feature (lost/found/adoption) made me feel connected to the community. \\
    Item 8 & The overall design feels different from typical adoption apps. \\
    \bottomrule
    \end{tabular}
\end{table}

\subsubsection{General Usability}

Across the four SUS-related items, the results show consistently high ratings, indicating strong usability of the prototype. Most respondents agreed that the system was easy to use (Item 1; M = 4.49, SD = 0.69), with 94.44\% selecting either 4 or 5. Confidence in using the prototype was similarly high (Item 2; M = 4.54, SD = 0.81), with 91.67\% of participants agreeing or strongly agreeing. Perceptions of how well the features worked together were even more positive (Item 3; M = 4.69, SD = 0.52), with 97.22\% indicating agreement. Finally, willingness to use the system in the future received the highest mean rating (Item 4; M = 4.74, SD = 0.61), with 91.67\% selecting 4 or 5. Taken together, these findings suggest that participants not only found the prototype usable and coherent but also expressed strong intent to adopt it if implemented. Table \ref{tab:general_usability} shows the summary of the results.

\begin{table}[h]
    \centering
    \caption{Descriptive statistics for general usability items (N = 35). See Table~\ref{tab:item_mapping} for item statements.}
    \label{tab:general_usability}
    \begin{tabular}{lrrr}
    \toprule
    \textbf{Item} & \textbf{Mean} & \textbf{SD} & \textbf{\% Agree (4--5)} \\
    \midrule
    Item 1 & 4.49 & 0.69 & 94.44 \\
    Item 2 & 4.54 & 0.81 & 91.67 \\
    Item 3 & 4.69 & 0.52 & 97.22 \\
    Item 4 & 4.74 & 0.61 & 91.67 \\
    \bottomrule
    \end{tabular}
\end{table}

\subsubsection{Feature-Specific Usability}

For the feature-specific items, Table \ref{tab:feature_usability} shows the summary of the results. Participants gave consistently high ratings across all major components of the prototype. The matchmaking feature successfully conveyed the intended sense of chance or serendipity (M = 4.71, SD = 0.61), with 91.67\% rating it positively (4 or 5). The proximity and location aspect was also well received (M = 4.51, SD = 0.69), as 88.89\% agreed it was useful in surfacing nearby adoption opportunities. The reporting feature emerged as one of the most strongly endorsed elements (M = 4.60, SD = 0.60), with 94.44\% providing positive ratings, reflecting its effectiveness in fostering a sense of community connection. Finally, the overall design of the system was perceived as distinctive compared to typical adoption apps (M = 4.69, SD = 0.52), with 97.22\% of respondents acknowledging its uniqueness. These results suggest that both the playful and practical aspects of the prototype were recognized and valued by participants.

\begin{table}[h]
    \centering
    \caption{Descriptive statistics for feature-specific usability items (N = 35). See Table~\ref{tab:item_mapping} for item statements.}
    \label{tab:feature_usability}
    \begin{tabular}{lrrr}
    \toprule
    \textbf{Item} & \textbf{Mean} & \textbf{SD} & \textbf{\% Agree (4--5)} \\
    \midrule
    Item 5 & 4.71 & 0.61 & 91.67 \\
    Item 6 & 4.51 & 0.69 & 88.89 \\
    Item 7 & 4.6 & 0.6 & 94.44 \\
    Item 8 & 4.69 & 0.52 & 97.22 \\
    \bottomrule
    \end{tabular}
\end{table}

\begin{figure}[htpb]
    \centering
    \includegraphics[width=\linewidth]{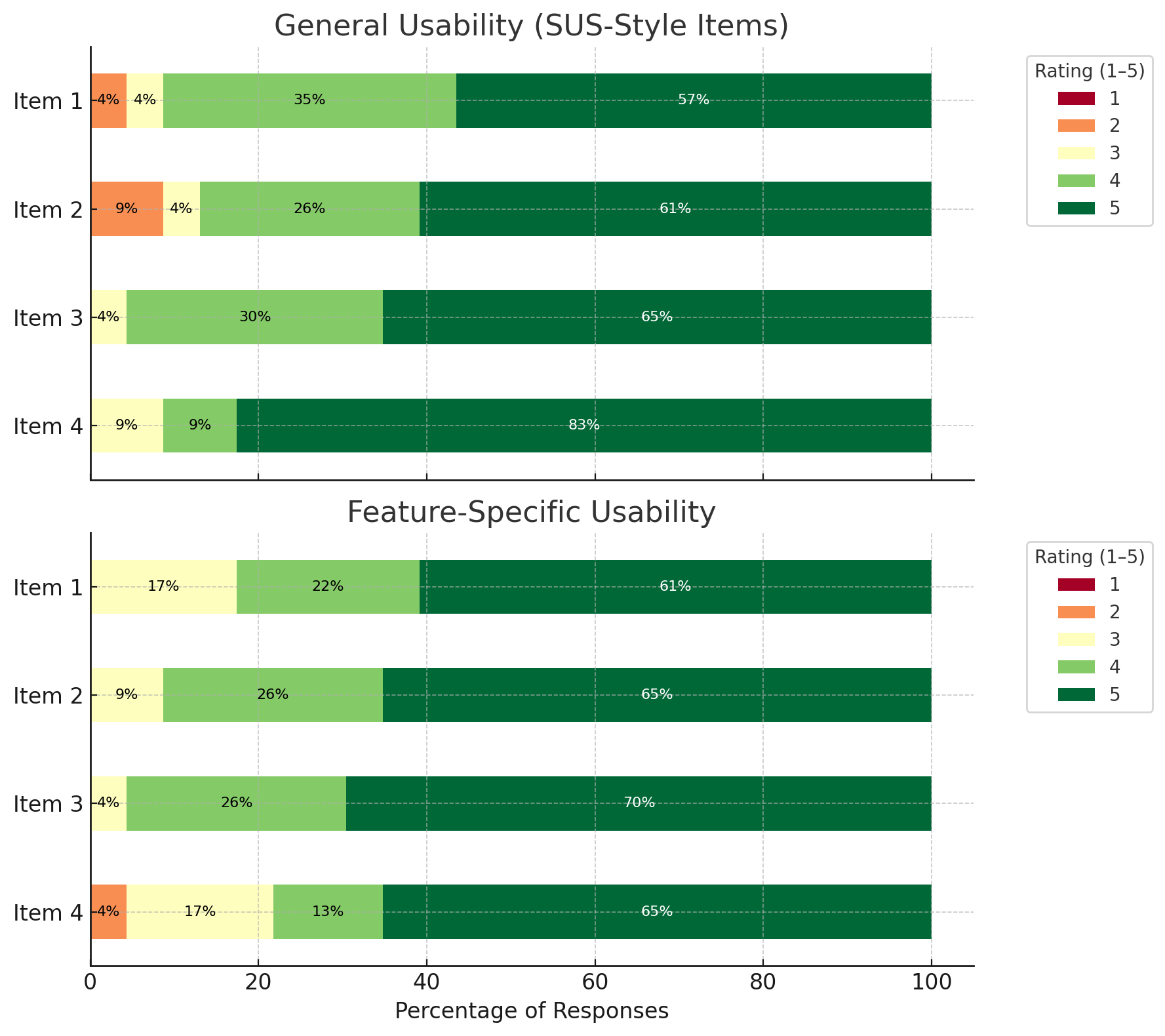}
    \caption{Distribution of responses across usability items. 
    Items are numbered (1--8) for readability; see Table~\ref{tab:item_mapping} for full questionnaire statements.}
    \label{fig:usability_likert}
    \Description{Two simple bar charts show how people rated the app’s usability on a 1–5 scale. Each row is a questionnaire item, and the bars are mostly dark on the right side, indicating many ratings of 4 and 5 and generally positive feedback.}

\end{figure}

Figure \ref{fig:usability_likert} presents the distribution of responses across all eight usability items, with item numbers mapped to their full statements in Table \ref{tab:item_mapping}. The results show a clear skew toward positive ratings (4–5) across both general usability (Items 1–4) and feature-specific evaluations (Items 5–8). In particular, Items 3, 4, and 7 received the strongest endorsements, indicating that participants found the prototype well integrated, expressed willingness to use it in the future, and valued the community-oriented reporting feature. Even the lowest-rated items (Items 2 and 8) still demonstrated a majority of agreement, suggesting overall consistency in favorable perceptions. Together, the figure and mapping highlight that the prototype was perceived as both usable and distinctive, with strong support for its key features.

\begin{table*}[!htpb]
\centering
\caption{Prototype task performance (Maze.co usability test)}
\label{tab:prototype-performance}
\begin{tabular}{@{} l  c  c  c  c @{}}
\toprule
Task & Success rate & Drop-off & Avg.\ duration (s) & Responses \\
\midrule
Login to App                      & 96.80\% & 3.20\%  & 82.0   & 31 \\
Exploring Matchmaking             & 92.30\% & 7.70\%  & 162.7  & 26 \\
Adopting a cat                    & 90.00\% & 10.00\% & 52.2   & 20 \\
Exploring Lost and Found          & 85.00\% & 15.00\% & 51.7   & 20 \\
Create a post --- Part 1          & 81.30\% & 18.80\% & 73.0   & 16 \\
Create a post --- Part 2          & 81.30\% & 18.80\% & 39.7   & 16 \\
\bottomrule
\end{tabular}
\end{table*}

The usability test, as shown in Table 6, shows that participants were generally successful in completing the main tasks within the prototype. These results are based on the 23 participants who completed the Maze-based usability tasks. Core features such as logging in, exploring matchmaking, and adopting a cat achieved high success rates above 90\%, suggesting that these functions were intuitive and easy to navigate. However, tasks related to content creation, particularly creating posts, recorded lower success rates of around 81.3\% with higher drop-off rates, indicating areas that may require further refinement or clearer guidance for users. Similarly, exploring the lost-and-found page had a 15\% drop-off, suggesting some usability challenges in accessing or interpreting this feature. Despite these variations, average completion times remained reasonable across all tasks. While the Maze report shows up to 31 responses for certain tasks, this inflation reflects duplicate entries when some participants refreshed the task state; the unique sample size for the Stage 1 usability test remains 23.

\subsection{Qualitative Findings}

In addition to the quantitative ratings, participants provided open-ended feedback regarding their experiences with the prototype. These responses were thematically analyzed to identify common perceptions, strengths, and suggestions for improvement.

\subsubsection{Positive Aspects Highlighted by Participants}

Responses to the question on what participants liked most about the prototype revealed several recurring themes. Many emphasized the system’s ease of use and simplicity, noting that the interface was straightforward and user-friendly. The location and proximity features, particularly the map function for identifying nearby or lost cats, were frequently highlighted as valuable and practical. Several participants also pointed to the matchmaking aspect, describing it as engaging, fun, and even exciting, which contributed to a more personal adoption experience. The lost-and-found functionality was similarly appreciated for its usefulness in tracking and reporting cats within the community. In addition, participants mentioned the prototype’s convenience and efficiency, characterizing it as helpful, time-saving, and designed to make adoption easier. A smaller number also commented on the visual design, specifically the color palette and graphics, which they found pleasant and not straining to the eyes.

In addition, several participants noted the prototype’s unique concept and perceived social impact. They felt that the system addressed gaps in existing platforms (e.g., generic Facebook posts), offered hope to owners searching for lost cats, and could raise awareness about responsible adoption within the community.

\subsubsection{User Suggestions for Improvement}

When asked about potential improvements, participants offered a range of suggestions alongside several indications of overall satisfaction. Many users pointed out the need for interface refinements, such as adding a clearer back button, ensuring consistency in the navigation bar, maintaining a uniform theme across pages, and clarifying the initial call-to-action so that first-time users know where to start. Others emphasized the value of guidance and support, suggesting simple guides that explain how the features work, outline adoption requirements, and offer basic care tips for first-time adopters.

Several responses focused on feature enhancements, including breed or personality filters, status indicators, and tools to reduce redundant postings, while some participants recommended improving the accuracy of the map feature. Participants also proposed strengthening trust and safety mechanisms by adding features such as filters or checks for potential scammers and more detailed cat profiles, including health histories, vaccination records, and deworming information. A few respondents suggested post-adoption support features and options to share ongoing updates through “stories” or “highlights,” which could help build a sense of continuity and community.

In terms of interaction and integration, some users recommended making the system more interactive, supporting multimedia uploads such as videos, and integrating with tools they already use to better fit the app into their daily routines. Others mentioned adding smoother onboarding options, such as biometric login (e.g., Face ID), and even exploring logistics-related features like facilitating long-distance transport with status updates. One participant also proposed extending the system to include other pets beyond cats. Notably, a considerable portion of respondents indicated that they would not change anything, reflecting a general sense of satisfaction with the current prototype design.

\subsubsection{Perceived Value of the CDS Approach}

The CDS-inspired concept was generally perceived positively by participants, who often described it as convenient and efficient because it reduces the need for active searching and instead allows cats to be “distributed” to potential adopters. The idea was also seen as playful and creative, as it was felt to capture the serendipitous ways cats often appear in people’s lives and to transform adoption into a sense of connection rather than a purely transactional process. The concept was further viewed as having strong potential for supporting adoption, welfare, and community building, with several remarks suggesting that more cats could find shelter and care through such a system. At the same time, some reservations were raised: questions were posed about how the approach would work for truly stray cats without clear owners, and it was noted that the design might be confusing for less tech-savvy users or for those who still prefer more direct control when browsing available cats.

\section{Discussion and Design Implications}

\subsection{Discussion}

This study contributes a novel methodological approach by leveraging netnography to systematically extract mental models embedded in internet culture. Specifically, by analyzing how the Cat Distribution System (CDS) narrative circulates and evolves on TikTok, we identified core mental models that shape how users conceptualize adoption - emphasizing serendipity, proximity, environment, and decision, as part of the Serendipitous Adoption Framework. We then translated these models into design principles for adoption platforms to demonstrate that aligning systems with users' existing cognitive schemas yields interfaces that feel intuitive and meaningful rather than imposed. This approach exemplifies how design can be anchored not in designer assumptions or abstract theory, but in the lived experiences and folk theories already circulating within target communities.

The evaluation shows that CDS-inspired adoption systems can combine practicality and playfulness in animal welfare technology. Participants gave high usability ratings, which suggests the prototype works well and that the design metaphor of cats being “distributed” instead of actively searched for resonates with users. This idea turns adoption into a more personal and unexpected experience. Several participants described it as exciting, creative, and similar to how cats often come into people’s lives by surprise.

The results also reveal some important design challenges. Users liked the automated matchmaking feature for being new and efficient, but some wanted more ways to browse and clearer information about how suggestions were made. Location and reporting features were seen as practical and good for building community, but there were concerns about accuracy, repeated information, and how consistent the interface was. These issues show that adoption platforms need to balance playfulness with reliability and give users enough control.

Another key finding is that the prototype presents adoption as a community activity, not just a personal decision. The reporting feature especially helped create a sense of shared responsibility, allowing adopters, owners, and rescuers to work together for cat welfare. This matches recent trends in HCI that encourage systems to improve both usability and social support. 

While the results provide promising insights into the usability and potential of the CDS-inspired prototype, several limitations remain. First, we tested the system with a relatively small convenience sample: 35 people interacted with the prototype, but only 23 completed the Stage 1 Maze-based usability tasks and questionnaire that supports the quantitative findings. This modest and demographically narrow group limits the generalizability of the results, and perspectives from a wider population of adopters, shelter workers, and community volunteers remain underexplored. Second, we evaluated a medium-fidelity prototype rather than a fully deployed system. Participants’ experiences may therefore differ from real-world adoption dynamics, particularly in terms of performance, reliability, and long-term engagement. It also may not capture real-world challenges, for instance, data accuracy in geolocation features and scalability. Third, although many participants responded positively to the CDS-inspired framing, concerns about user control and accessibility point to the need for deeper investigation into how different audiences interpret and interact with the concept. Not all participants were familiar with the idea of CDS, so some may have understood or valued the metaphor differently.

Future work should address these gaps by testing the prototype with larger and more diverse groups, including actual adopters and animal welfare organizations. Longitudinal studies could provide insights into sustained use, adoption outcomes, and community impact over time. Additionally, expanding the prototype to support other animals beyond cats, integrating health and welfare records, and refining matchmaking algorithms to balance serendipity with user choice could enhance both utility and trust.

\subsection{Design Implications}

These insights inform a set of design implications that show how future systems can preserve the strengths of the CDS approach while addressing its limitations:

\subsubsection{Design for Serendipity with Agency}
The CDS-inspired model demonstrates how playful metaphors can make adoption feel personal and engaging. However, feedback revealed that users also want the ability to browse or filter beyond automated matches. Adoption platforms should preserve serendipity while layering in user control, ensuring that choice and transparency coexist with chance.

\subsubsection{Embed Community as a Core Function}
The strong reception of the reporting feature indicates that adoption is not just an individual act but a community process. Future designs should treat community reporting, shared visibility of lost/found cases, and collaborative welfare as first-class elements, not add-ons, thereby positioning platforms as civic tools as much as adoption services.

\subsubsection{Use Location Features Responsibly}
Proximity-based matching was valued for practicality, but concerns about map accuracy and redundant reports highlight the need for careful handling of geolocation data. Systems should incorporate verification mechanisms (e.g., status indicators, duplicate detection) to maintain trust and reduce noise, ensuring that location aids adoption rather than creating confusion.

\subsubsection{Lower Barriers Through Guidance and Education}
Requests for guides, adoption tips, and health records emphasize the importance of supporting first-time adopters. Embedding contextual help and welfare information directly into the flow of adoption can build confidence and support responsible pet ownership.

\subsubsection{Balance Playfulness with Accessibility}
The CDS metaphor was commended for being creative and fun, but some participants worried it might confuse less tech-savvy users. Designers should ensure that playful elements do not obscure clarity, and should test across diverse demographics to calibrate metaphors to different levels of digital literacy.

\section{Conclusion}
This study explored how the meme about the Cat Distribution System (CDS) can inform the design methods for adoption platforms. From the mental model derived from online artifacts, we were able to reframe the adoption process from the lived experiences of the users. The prototype resulting from the Serendipitous Adoption Framework introduced a playful yet practical model that integrates matchmaking, location awareness, and community reporting. Quantitative findings showed consistently high usability ratings, while qualitative feedback highlighted strengths in simplicity, convenience, and community orientation, alongside constructive suggestions for improvement.

The results suggest that CDS-inspired design not only resonates with users but also opens new pathways for thinking about adoption as a blend of serendipity, agency, and collective responsibility. By balancing playfulness with reliability, embedding community features, and supporting adopter guidance, future systems can extend the promise of CDS toward more effective and humane adoption practices. Ultimately, this work demonstrates how culturally resonant metaphors, even those rooted in internet folklore, can inspire novel approaches to designing technologies for social good.


\begin{acks}
C.A.G. Angcana and J.I.N. Gojo Cruz gratefully acknowledges the support provided by the University of the Philippines Los Baños through its Academic Development Fund.
\end{acks}

\bibliographystyle{ACM-Reference-Format}
\bibliography{revised}

\end{document}